\documentclass[preprint,prb,groupedaddress,showpacs]{revtex4}
\usepackage{epsfig}
\begin{document}
\title{Charge decoherence in laterally coupled quantum dots due to
electron-phonon interactions}
\author{V.N. Stavrou}
\author{Xuedong Hu}
\affiliation{Department of Physics, State University of New York at Buffalo,
New York 14260, USA}
\begin{abstract}
We investigate electron charge decoherence in a laterally-coupled
single-electron semiconductor double quantum dot through electron-phonon
interaction.  We analytically and numerically evaluate the relaxation and
dephasing rates due to electron coupling to both acoustic and optical
phonons, and explore the system parameter space in terms of interdot
distance, strength of single-dot confinement, and inter-dot coupling
strength.  Our numerical results show that the electron scattering rates are
strongly dependent on the strength of the electron confinement and the size
of the system.  In addition, although the most dominant factor that determines
the charge decoherence rate is the energy splitting between the charge qubit
states, the details of the double dot configuration is also very important.
\end{abstract}
\maketitle
\date{\today}

\section{INTRODUCTION}

Ever since Peter Shor showed that a purely quantum mechanical computer can be
used to achieve exponential speedup (compared to classical computers) in
solving the prime factoring problem,\cite{Shor} there has been widespread
interests in the study of quantum information science in general, and in
building a practical quantum computer in particular.\cite{Reviews,Nielsen} 
Some of the most prominent proposals are based on solid state structures,
whether superconducting nanocircuits\cite{Makhlin,Bouchiat,Nakamura,Chiorescu}
or semiconductor nanostructures such as quantum dots and regular donor
arrays.\cite{LD,Kane,Tanamoto,Sherwin,Fed,Hollenberg}  A major presumed
advantage of these systems, especially the semiconductor artificial
structures, are their potential scalability, supported by the powerful and
advanced semiconductor industry.

The great interest in semiconductor-based quantum computer architectures has
prompted extensive studies of a variety of physical properties of
nanostructures that are relevant to single electron spin or charge degree of
freedom.\cite{Hu}  In the present paper we focus on the decoherence properties
of a charge qubit in semiconductors.\cite{Tanamoto,Fed,Hollenberg}  For a
charge qubit based on a single electron in a semiconductor double quantum dot
(QD), charge decoherence has two important channels: Coulomb interaction to
the background charge fluctuation and electron-phonon
interaction.\cite{Hayashi,Petta,Fujisawa,Barrett,Fedichkin,Voroj,HKD}  The former
decoherence channel is widely present in all nanostructures.  Presumably it
is due to coupling with defects in the system and is thus extrinsic.  The
latter, however, is intrinsic to any solid state host material for the charge
qubit.

In the following we present our study of charge qubit decoherence caused by
electron-phonon interaction in a horizontally coupled two-dimensional GaAs
double quantum dot.  In Section II, we identify the electronic states we are
interested in, and clarify the relevant types of phonons and electron-phonon
interactions involved.  We then derive the various relaxation and dephasing
rates for double-quantum-dot-trapped single electrons.  In Section III and IV, 
we show our results on charge relaxation and dephasing rates for a variety of 
configurations and states, and discuss the physical pictures and implications. 
Section V presents a summary of our results and our
conclusions.

%************************************************
\section{Theoretical description of electron-phonon interaction}
%************************************************
\subsection{A single confined electron in two coupled quantum dots}
%************************************************

In this study we consider gated lateral quantum dots in a AlGaAs/GaAs/AlGaAs
quantum well (QW).  The growth direction ($z$ direction, or vertical
direction) confinement is due to the higher bandgap of the barrier material
of AlGaAs.  The lateral confinement is produced by the electrostatic
potential from surface metallic gates.  In general, the vertical direction
confinement length ($\sim 10$ nm) is much smaller than the lateral
confinement length ($50$ nm), so that we can safely treat the dynamics
along vertical and horizontal directions as decoupled.  The system
Hamiltonian (within the effective mass and envelope function approximations)
is thus
\begin{equation}
\label{hamil}
\hat{\mathcal{H}} = \hat{\mathcal{H}}_{\parallel} + \hat{\mathcal{H}_{z}} \,,
\end{equation}
where the growth direction component $\hat{\mathcal{H}_{z}}$ takes the form 
\begin{equation}
\label{hamil_z}
\hat{\mathcal{H}_{z}} =
-\frac{\hbar}{2}\partial_{z}\frac{1}{m^{\ast}(z)}\partial_{z}
                        +
V_{0}\Theta\left(\left|z\right|-L_{z}\right) \,.
\end{equation}
Here $m^{\ast}(z)$ is the electron effective mass and $V_{0}$ the offset
between the band edges of the GaAs well and the AlGaAs barrier.
For simplicity, we take the $z$-direction wavefunction as the
wavefunction of an infinite QW $(V_{0} \rightarrow \infty)$.  In this work, we
do not consider excitations along $z$-direction because of the much higher
excitation energy (compared to the lateral direction), so that the
$z$-direction wavefunction is always given by $\psi_{\it{z}}
\left(\it{z}\right) = \mathcal{A} \cos\left(\pi z/2L_{z} \right)$ where
$\mathcal{A}$ is a coefficient to be determined by normalization and $2L_{z}$
is the width of the QW. 

The lateral confinement is assumed to be parabolic for a single QD, so that a
single electron Hamiltonian in the lateral direction is
\begin {equation}
\label{hamil_xy}
\hat{\mathcal{H}}_{\parallel} = 
-\frac{\hbar^{2}}{2m^{\ast}}\nabla^{2}
+\frac{1}{2}m^{\ast}\omega^{2}_{0}r_{\parallel}^{2}
\end {equation}
where $\omega_{0}$ describes the strength of the harmonic confinement in the
$x-y$ plane.  The total electron wavefunction can now be written as a product
of 
\begin {equation}
\label{envelope}
\psi(\bf{r}) = \psi_{\parallel}\left(\bf{r}_{\parallel}\right)
               \psi_{\it{z}}\left(\it{z}\right)
\end {equation}

In the case of a single QD, the two-dimensional (2D) one-electron
wavefunctions are essentially 2D harmonic oscillator functions
\cite{Jacak,Bockel} and are described in terms of the principal quantum
number ${\it{n}}={0,1,2,...}$ and the angular momentum quantum number
${\it{m}}={0,\pm1,\pm2,...}$ as
\begin {equation}
\label{psi_xy}
\psi_{\parallel}^{(n,m)}\left(\tilde{\rho}, \theta \right) = 
\sqrt{\frac{{n}!}{\pi l^{2}\left({n}+\left|{\it{m}}\right|\right)!}}
\tilde{\rho}^{\left|{\it{m}}\right|}e^{-\tilde{\rho}^{2}/2}
e^{{\it{im}\theta}}
{\mathcal{L}}_{{n}}^{\left|{\it{m}}\right|}\left(\tilde{\rho}^{2}\right)
\end {equation}
where ${\mathcal{L}}_{n}^{\left|{\it{m}}\right|}\left(\tilde{\rho}^{2}\right)$
are the Laguerre polynomials, and $\tilde{\rho} = |{\bf r}_{\parallel}|/{\it
l}$ is a scaled radius, with ${\it l}=\sqrt{\hbar/m^{\ast}\omega_{0}}$.  The
corresponding eigenvalues are
\begin {equation}
E_{nm} = \left(2n+\left|{\it{m}}\right|+1\right)\hbar\omega_{0} \,.
\end {equation} 

For two QDs that are horizontally coupled, we use a simple in-plane
confinement of two parabolic wells separated by an inter-dot distance
$2\alpha$:
\begin {equation}
\label{Vc}
V_{c} = \frac{1}{2}m^{\ast}\omega_{0}^{2} \ {\rm min} \{
\left(x-\alpha\right)^{2}+y^{2},~\ \left(x+\alpha\right)^{2}+y^{2} \}
\end {equation}
The single electron wavefunction for the lateral direction is, in general,
given by a superposition of the single-dot wavefunctions: 
\begin {equation}
\label{superposition}
\left |\Psi_{\|}\right> = \sum_{k}{C_{k}\left |\psi_{\|,L}^{k}\right> + 
                              D_{k}\left |\psi_{\|,R}^{k}\right>} \,,
\end {equation}
and the total wavefunction of the system of the coupled QDs is
\begin {equation}
\label{wavefunction}
\Psi(\bf{r}) = \Psi_{\|}\left(\bf{r_{\|}}\right)
                \psi_{\it{z}}\left(\it{z}\right)
\end {equation}
Notice that for charge qubits there is only a single electron in a double dot,
in contrast with spin qubits, where each quantum dot has an electron and
double dot is only for two-qubit operations.\cite{LD,HD}  In the present
study, the wavefunctions for the coupled-QD are calculated numerically by
direct diagonalization, using reasonable parameters of a GaAs QW.

\subsection{Electron-Acoustic phonons coupling}

In a polar semiconductor like GaAs, electrons couple to all types of phonons. 
More specifically, in GaAs electrons couple to longitudinal acoustic phonons
through a deformation potential, to longitudinal and transverse acoustic
phonons through piezoelectric interaction, and to the optical phonons through
the polar interaction.\cite{Mahan} The deformation potential electron-phonon 
interaction is given by 
\begin{equation}
\label{H_D}
H_{D} = D \sum_{\bf q} \left( \frac{\hbar}{2 \rho_m V \omega_{\bf q}}
\right)^{1/2} |{\bf q}| \rho({\bf q}) (a_{\bf q}+a_{-\bf q}^{\dagger}) \,,
\end{equation}
where $D$ is the deformation constant, $\rho_m$ is the mass density of the
host material, $V$ is the volume of the sample, $a_{\bf q}$ and $a_{-\bf
q}^\dagger$ are phonon annihilation and creation operators, and $\rho({\bf
q})$ is the electron density operator.  Table I presents the material
parameters used in our numerical calculation.

Electrons can interact also with longitudinal and transverse acoustic phonons
through piezoelectric interaction.  This type of interaction is essentially
due to the lack of symmetry of in the crystal, thus for materials like Si,
which has crystal inversion symmetry, the piezoelectric interaction is not
present.  On the other hand, the crystal of GaAs lacks inversion symmetry, so
piezoelectric interaction is nonvanishing.  The electric displacement $\cal
D$ is related to the electric field $\cal E$, strain $S$, and the permitivity
tensor $\epsilon$ in a piezoelectric crystal by \cite{Ridley82}
\begin {equation}
{\cal{D}}_{i} = \sum_{i}{\epsilon_{ij}{\cal{E}}_{j}}+\sum_{k,l}{e_{ikl}S_{kl}}
\end {equation}
where the third rank tensor $e_{ikl}$ is the piezoelectric constant tensor. 
As a result, the electron-phonon coupling due to the piezoelectric effect
is\cite{Mahan}
\begin{equation}
H_{P} = {\it{i}} \sum_{\bf q} \left( \frac{\hbar}{2 \rho_m V \omega_{\bf q}}
\right)^{1/2} \mathcal{M}^{pz}_{\lambda} (\hat{q}) \rho({\bf q}) (a_{\bf
q}+a_{-\bf
q}^{\dagger})
\end{equation}
where $\lambda$ denotes polarization of the acoustic phonons.  In the case of
zincblende crystals there is only one independent and nonvanishing
piezoelectric constant: $e_{14}=e_{25}=e_{36}$, so that the matrix element
${\mathcal {M}}^{pz}_{\lambda}$ is given by\cite{Mahan, MahanPAP1, Bruus1993}
\begin {equation}
\label{PZ}
{\mathcal{M}}^{pz}_{\lambda} (\hat{\bf q}) = 2 e \ e_{14}\left( \hat{q}_{x}
\hat{q}_{y} \xi_{z} + \hat{q}_{y} \hat{q}_{z} \xi_{x} + \hat{q}_{x}
\hat{q}_{z} \xi_{y} \right)
\end {equation}
where $\xi$ denotes the unit polarization vector and $e$ is the electron
charge.

Notice that the Hamiltonian for deformation potential and piezoelectric
interaction are real and imaginary, respectively, which allow us to
investigate separately these interactions and calculate the total
contribution by simply adding up the two rates.\cite{Mahan}    
\begin{table}[t]
\begin{tabular}{|c|c|c|c|c|c|c|c|c|c|}
\hline
 & $m/m_{e}$  & $D~(eV)$ & $c_{s}~(m/s)$ & $\rho~(Kgr/m^{3})$
    & $e_{14}~(V/m)$ & $\epsilon_{s}$ & $\epsilon_{\infty}$
    & $\omega_{LO}(meV)$ & $\omega_{TO}(meV)$\\
\hline
GaAs & 0.067  & 8.6 & 3700 & 5300 & 1.38$\times 10^{9}$
     & 12.9 & 10.89 & 36.25 & 33.29\\
\hline                                                                                    
\end{tabular}
\caption{Material parameters ($m_{e}$ is the electron mass).}
\end{table}

\subsection{Electron-Optical phonons coupling}

Even though the electronic energy involved in the study of charge qubits is
generally quite small (a few meV) compared to the optical phonon energy
($\sim$ 36 meV in GaAs), we will demonstrate later that
electron-optical-phonon interaction does play a role in the decoherence of
electron orbital states.  For the purpose of this calculation, we will use
the simple polar interaction in the bulk, neglecting the more intricate
details involving heterostructures.\cite{Essex,Ridley96,Stavrou03}  The
electron-phonon
interaction due to LO phonons is thus given by \cite{Mahan}
\begin{equation}
H_{OP} = \sum_{\bf q} \frac{M}{q\sqrt{V}}
\rho({\bf q}) (a_{\bf q}+a_{-\bf q}^{\dagger})
\end{equation}
and
\begin {equation}
\label{Coupling2}
M^{2} = 2\pi e^{2}\hbar\omega_{LO}
        \left( \frac{1}{\epsilon_\infty{}} -
\frac{1}{\epsilon_{s}} \right)    
\end {equation}
in which $\omega_{LO}$ is the longitudinal optical frequencies,
$\epsilon_{s}$ and $\epsilon_{\infty}$ are the static and high frequency
dielectric constant.

Now that we have identified all the electron-phonon interaction involved in
our system, we are ready to study what concrete forms they take in a double
quantum dot system, and evaluate the dephasing and relaxation rates 
for an electron in a double quantum dot due to its interaction with the
phonons.

%************************************************************************
\section{Relaxation and dephasing rates due to electron-phonon interaction}
%************************************************************************

\subsection{Electron-phonon coupling in a double quantum dot}

Before performing numerical evaluations of the electron charge relaxation and
dephasing rates, we first clarify the physical picture of charge decoherence
in a double quantum dot.  As an example and without loss of generality, let
us examine the deformation potential electron-phonon interaction given by
Eq.~(\ref{H_D}) in which $\rho({\bf q})$ is the Fourier transform of the
electron density operator: 
\begin{equation}
\rho({\bf q}) = \sum_{\kappa, \eta} c_{\kappa}^\dagger c_{\eta} \int d{\bf
r} \ e^{-i{\bf q} \cdot {\bf r}} \phi_{\kappa}^* ({\bf r}) \phi_{\eta} ({\bf
r}) \,,
\end{equation}
where $\kappa$ and $\eta$ are indices of electronic states, $c_\kappa$
and $c_\kappa^\dagger$ are electronic annihilation and creation operators
for the $\kappa$-state, while $\phi$ are the electron wavefunctions.  In the
context of a coupled double QD, we can choose the double dot eigenstates as
the basis for the single electron.  For the two lowest-energy double-dot
states $\Phi_\pm$, which are chosen as the charge qubit basis states,
$\kappa$ and $\eta$ take the values of $+$ and $-$ (from now on, these two
states will be equivalently referred to as the ground and first excited
states of the double dot, or the charge qubit states).  The electron-phonon
coupling Hamiltonian can then be conveniently written in this quasi-two-level
basis in terms of the Pauli spin matrices $\sigma_x$ and $\sigma_z$ (where
spin up and down states refer to the two electronic eigenstates):
\begin{eqnarray}
H_{D} & = & D \sum_{\bf q} \left( \frac{\hbar}{2 \rho_m V \omega_{\bf q}}
\right)^{1/2} |{\bf q}| \left(A_r ({\bf q}) \sigma_x + A_{\varphi} ({\bf q})
\sigma_z \right) \left( a_{\bf q}+a_{-\bf q}^{\dagger} \right) \,, \nonumber
\\
A_r ({\bf q}) & = & \langle -|e^{i{\bf q}\cdot{\bf r}}|+\rangle \,, \nonumber
\\
A_{\varphi} ({\bf q}) & = & \frac{1}{2} \left( \langle +|e^{i{\bf q}\cdot{\bf
r}}|+\rangle - \langle -|e^{i{\bf q}\cdot{\bf r}}|-\rangle \right) \,.
\end{eqnarray}

Since the basis for the quasi-two-level system are electron eigenstates for
the double dot, the term proportional to $\sigma_x$ above leads to transition
between the two electronic eigenstates and causes relaxation.  On the other
hand, the term proportional to $\sigma_z$ does not mix the electronic states,
so that it only causes fluctuations in the energy splitting between the two
electronic levels.  Therefore it only leads to pure dephasing between the two
electronic charge states, but not to relaxation.

For a discussion of the qualitative behavior of the electron-phonon coupling
in a double dot, we first analyze the simple situation where the two dots are
well separated and not strongly biased, so that only the two single-dot
ground orbital states are involved.  The relevant single-electron double-dot
states are then
\begin{equation}
\Phi_+ = a\phi_A({\bf r}) + b\phi_B({\bf r})~;~  \Phi_-= b\phi_A({\bf r}) -
a\phi_B({\bf r}), 
\label{eq:gaas1}
\end{equation}
with $\phi_{A(B)}({\bf r}) = \varphi({\bf r} - {\bf R}_{A(B)})\, u_0({\bf
r})$, where $\varphi({\bf r})$ is a slowly varying envelope function, and the
Bloch function at the conduction band minimum (${\bf k}=0$ at $\Gamma$ point)
is equal to the periodic part $u_0({\bf r})$.  Though we have chosen the
envelopes $\varphi$ centered at each well to be identical, they could as well
be different, as is generally the case for quantum dots.  For small energy
splittings between the $\Phi_\pm$ states, the fast oscillatory Bloch function
$u_{0} ({\bf r})$ can be integrated separately, so that the matrix element
$A_r$ can be written as\cite{HKD}:
\begin{eqnarray}
A_r ({\bf q}) & = & (ab^* - a^*b e^{i {\bf q} \cdot {\bf R}}) \int d{\bf r} ~
e^{i{\bf q} \cdot {\bf r}} [\varphi ({\bf r})]^2 \nonumber \\
& & + (|b|^2 - |a|^2) \int d{\bf r} ~ e^{i{\bf q} \cdot {\bf r}} \varphi ({\bf
r}) \varphi ({\bf r-R}) \,.
\label{eq:A_r}
\end{eqnarray}
Here the first integral is an on-site contribution modified by the phase
difference $e^{i {\bf q}\cdot {\bf R}}$ between the two dots, while the
second integral is a two-dot contribution that is generally much smaller
because of the small interdot overlap.

The dephasing matrix element $A_{\varphi}$ can be similarly calculated and the
result is\cite{HKD}
\begin{eqnarray}
A_{\varphi} ({\bf q}) & = & i \left(|b|^2 - |a|^2\right) e^{i{\bf q} \cdot
{\bf R}/2} \sin \frac{{\bf q} \cdot {\bf R}}{2} \int d{\bf r} ~ e^{i{\bf q}
\cdot {\bf r}} [\varphi ({\bf r})]^2 \nonumber \\
& & + (a^* b + ab^*) \int d{\bf r} ~ e^{i{\bf q} \cdot {\bf r}} \varphi ({\bf
r}) \varphi ({\bf r-R}) \,.
\label{eq:A_ph}
\end{eqnarray}
Here the prefactors $|b|^2 - |a|^2$ and $a^* b + ab^*$ are for intradot and
interdot integrals, exactly the opposite to those in Eq.~(\ref{eq:A_r}).  

Equations~\ref{eq:A_r} and \ref{eq:A_ph} clearly demonstrates that
electron-phonon interaction induced electron charge decoherence
is dominated by relaxation when $|b| \sim |a|$ so that $A_{\varphi}$ is
small, \cite{Barrett,Fedichkin} and by pure dephasing when $|b|$ and $|a|$ are
very different (so that, for example, $|b| \sim 1$ and $|a| \sim
0$).\cite{Fedichkin}  In other words, if the energy levels of the two quantum
dots are close to resonance, relaxation matrix element is much larger than
pure dephasing matrix element; while when the two single-dot levels are
biased, relaxation is suppressed.  These qualitative trends will persist even
when we consider more realistic electron eigenstates as studied in section
II.A.  The argument here is based on the assumption that the off-site
contributions are small because of a small wavefunction overlap between the
two dots.  Obviously, if the overlap is larger, the above trend becomes
weaker.

\subsection{Relaxation rates}

The electron relaxation rates associated with phonon emission (or absorption)
can be evaluated using Fermi's golden rule:
\begin {eqnarray}
\label{fermi}
\Gamma  &=&  \frac{2\pi}{\hbar}
\sum_{\bf{q}} 
\left| \left< \Psi^{(F)}({\bf{r}})\left|H^{\it{int}}
\right| \Psi^{(I)}({\bf{r}}) \right> \right|^{2}  
\delta\left(E_{F}-E_{I} \pm E_{\bf{q}} \right)
\nonumber\\ & &
\left(N_{B}(E_{\bf{q}},T_{lat})+\frac{1}{2} \pm \frac{1}{2}   \right)
\end {eqnarray}
Here labels 'I' and 'F' refer to the initial and final electron orbital states
respectively, the plus (minus) sign denotes emission (absorption) of a phonon,
and $N_{B}$ is the Bose-Einstein distribution for phonons with lattice
temperature $T_{\rm lat}$ (our calculations presented in this paper are all
done at $T_{lat}=0$, when phonon absorption can be neglected.).

Due to their large energies ($\sim 36$ meV for LO phonons in GaAs), optical
phonons do not contribute to electron orbital relaxation in a quantum dot
except for the highly excited states, which are irrelevant for the context of
charge-based quantum computing.  Therefore, the only phonons that contribute
to electron relaxation here are acoustic phonons.\cite{bulk_phonons}

\subsection{Pure dephasing rates}

Relaxation is not the only way charge qubits can be decohered.  If the energy
difference between the two charge states fluctuates, phase information will
get lost and decoherence occurs.  Such pure dephasing (in the sense that no
transition occurs between the two charge states) due to a bosonic bath has
been calculated before.\cite{Palma96,Duan}  Pure dephasing due to
electron-acoustic-phonon interaction has also been evaluated for spherical
quantum dots and donors in semiconductors.\cite{Fedichkin}  The density
operator of an electron in a boson bath can be written in a general
expression as
\begin{eqnarray} 
\label{density_matrix}
\rho(t) = \left( \begin{array}{cc} \rho_{00}(0) & \rho_{01}(0)e^{-B^{2}(\Delta
t)+i\varepsilon \Delta t/\hbar} \\ 
\rho_{10}(0)e^{-B^{2}(\Delta t)-i\varepsilon \Delta t/\hbar} & \rho_{11}(0)
\end{array} \right)
\end{eqnarray} 
where $\varepsilon$ is the energy splitting between the electron energy
levels.  In short, pure dephasing cause a decay in the off-diagonal element of
the density matrix for the two-level system that makes up the charge qubit
\cite{Palma96,Duan,Fedichkin}:
\begin {equation}
\rho_{01}(t)\sim \rho_{01}(0)e^{-B^{2}(t)} \,,
\label{eq:rho_10}
\end {equation}
where the exponent function $B^{2}(t)$ is defined by
\begin {equation}
\label{spectral}
B^{2}(t) = \frac{V}{\hbar^{2} \pi^{3}} \int{d^{3}{\bf q} \frac{|g({\bf
q})|^{2}}{\omega_{\bf q}^2} \sin^{2} \frac{\omega_{\bf q} t}{2} \coth
\frac{\hbar \omega_{\bf q}}{2k_{B} T}} \,.
\end {equation}
Here $\omega_{\bf q}$ is the frequency of the phonons.  In our study, we have
investigated dephasing effects due to both acoustic and optical phonons.  For
acoustic phonons, we choose $\omega_{\bf q} = qc_{s}$ for the relevant
branches, while for longitudinal optical phonons, we choose $\omega_{\bf q} =
\omega_{LO}$.  It can be shown straightforwardly that the phonons that
contribute significantly to pure dephasing are zone-center phonons (small
$|{\bf q}|$ values), so that choosing linear and constant dispersion for
acoustic and optical phonons is an excellent approximation.  The coupling
constants $g({\bf q})$ due to deformation potential, piezoelectric and optical
phonons are respectively given by 
\begin {eqnarray}
\label{dephas_Elem_D}
g_{\rm def} ({\bf q}) = D \sqrt{\frac{\hbar q}{2\rho c_{s}V}}
\mathcal{I}({\bf q}) \,, \\
\label{dephas_Elem_P}
g_{\rm piezo} ({\bf q}) = {\mathcal{M}}^{pz}_{\lambda}({\bf q}) \ 
\sqrt{\frac{\hbar}{2\rho c_{s}V}}
\mathcal{I}({\bf q}) \,, \\
\label{dephas_Elem_LO}
g_{\rm polar}({\bf q}) = \frac{M}{q\sqrt{V}}
                             \mathcal{I}({\bf q}) \,,
\end {eqnarray}
where $\mathcal{I}({\bf q})$ is defined by
\begin {eqnarray}
\mathcal{I}({\bf q}) = \frac{1}{2}\left(\left<
\Psi^{-}({\bf{r}})\left|e^{\mp{\it{i}}{\bf{q \cdot r}}}
\right| \Psi^{-}({\bf{r}})\right>
- \left< \Psi^{+}({\bf{r}})\left|e^{\mp{\it{i}}{\bf{q \cdot r}}}
\right| \Psi^{+}({\bf{r}})\right>
\right) = - A_{\varphi} ({\bf q}) \,.
\end {eqnarray}
Here $\pm$ refer to the two states for the double dot charge qubit.  Notice
that all the integrals in this study are carried out using the Monte-Carlo
technique.

%*********************************
\section{RESULTS AND DISCUSSIONS}
%*********************************
%
\subsection{Charge relaxation due to acoustic phonons}

We first calculate the electron relaxation rate from the excited charge qubit
state (the first excited state of the double quantum dot) and explore its
behavior as a function of the double dot parameters such as interdot distance
and strength of the single dot confinement.  Notice that throughout this
paper, the QW width takes on a fixed value of $2L_{z} = 6$ nm.  We have also
done calculations for a well width of 10 nm and the results are only slightly
different. 

\begin{figure}[ht]
\begin{center}
\includegraphics[trim= 0 0 0 -45, width=3.0in]{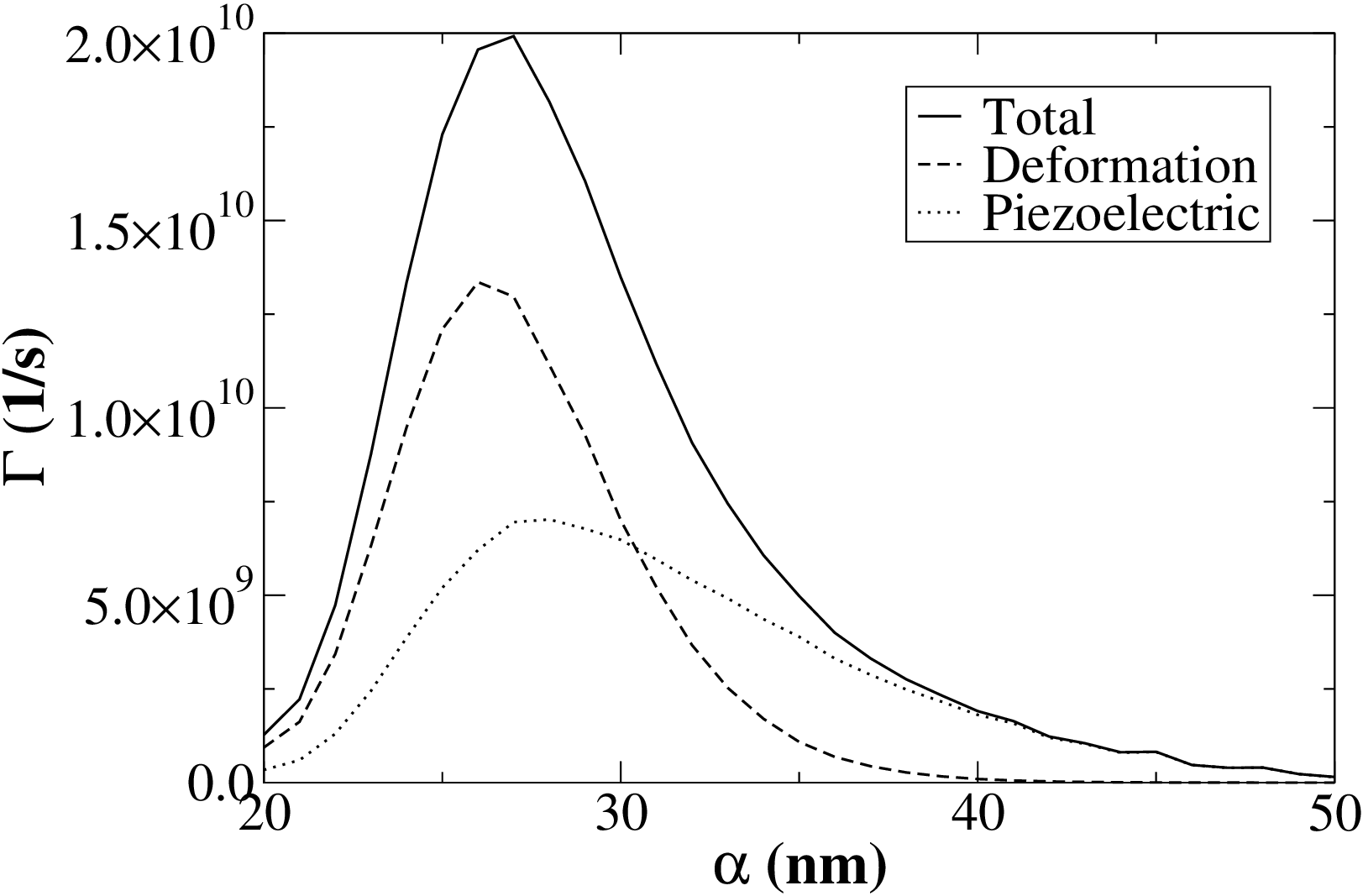}
\protect\caption{Relaxation rates of an electron in the first excited state
through acoustic phonon emission as a function of the half interdot distance
$\alpha$.  The relaxation rates due to deformation potential interaction,
piezoelectric interaction, and the total relaxation rates are presented by
dashed, dotted, and straight lines respectively.  The strength of the lateral
confinement is $\hbar\omega_{0}=3$ meV.}
\label{fig_1} 
\end{center}
\end{figure}

Figure \ref{fig_1} shows the electron relaxation rate as a function of the
interdot distance for an electron that is initially in the first excited
state.  The relaxation process is dominated by the emission of one acoustic
phonon.  For small inter-dot separations, the contribution due to deformation
potential interaction is larger than due to piezoelectric interaction.  For
larger inter-dot separation, however, piezoelectric coupling becomes the
dominant contributor because of the different wavevector dependence in the
deformation and piezoelectric matrix elements ($\sqrt{q}$ for deformation
potential versus $1/\sqrt{q}$ for piezoelectric interaction).  Since phonon
density of state goes to zero at small energy $\propto E^2$, the relaxation
rate decreases with decreasing energy splitting between the initial and final
states, which is the case for both very shallow confinement wells and for
largely separated quantum dots.  Experimentally, it has been found
\cite{Hayashi} that for large dots and large interdot distance ($\sim 150$
nm) relaxation rates due to electron-phonon coupling should be smaller than
$10^{9}~s^{-1}$.  Our calculations are quite consistent with these
experimental observations.

\begin{figure}[ht]
\begin{center}
\includegraphics[trim= 0 0 0 -45, width=3.0in]{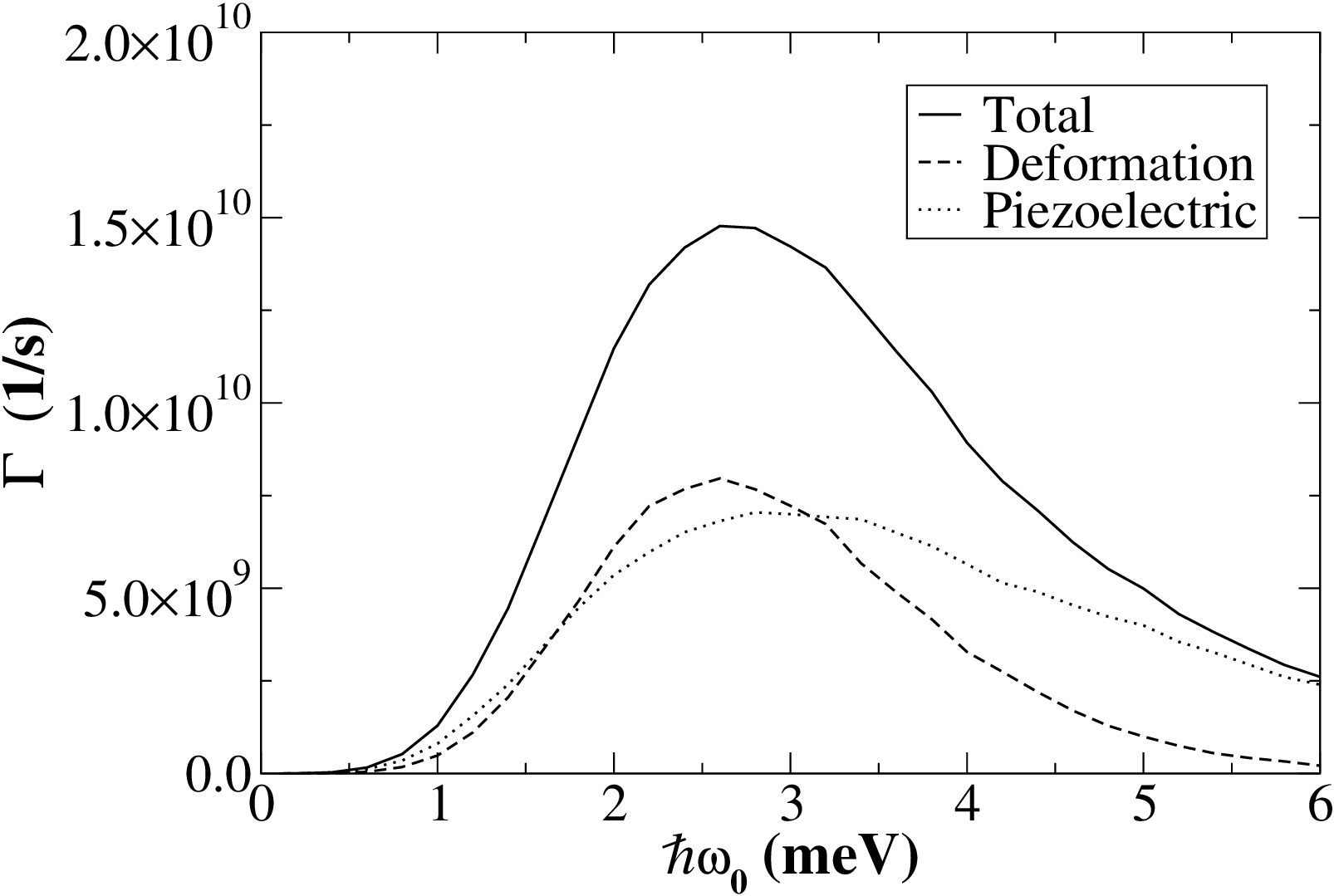}
\end{center}
\protect\caption{The relaxation rates versus the strength of the confinement.
The scattering rates due to deformation potential, piezoelectric phonons and
total relaxation rates are presented by dashed, dotted and straight line
respectively.  The interdot distance is $2\alpha = 60$ nm.}
\label{fig_2}
\end{figure}

\begin{figure}[ht]
\begin{center}
\includegraphics[trim= 0 0 0 -45, width=3.0in]{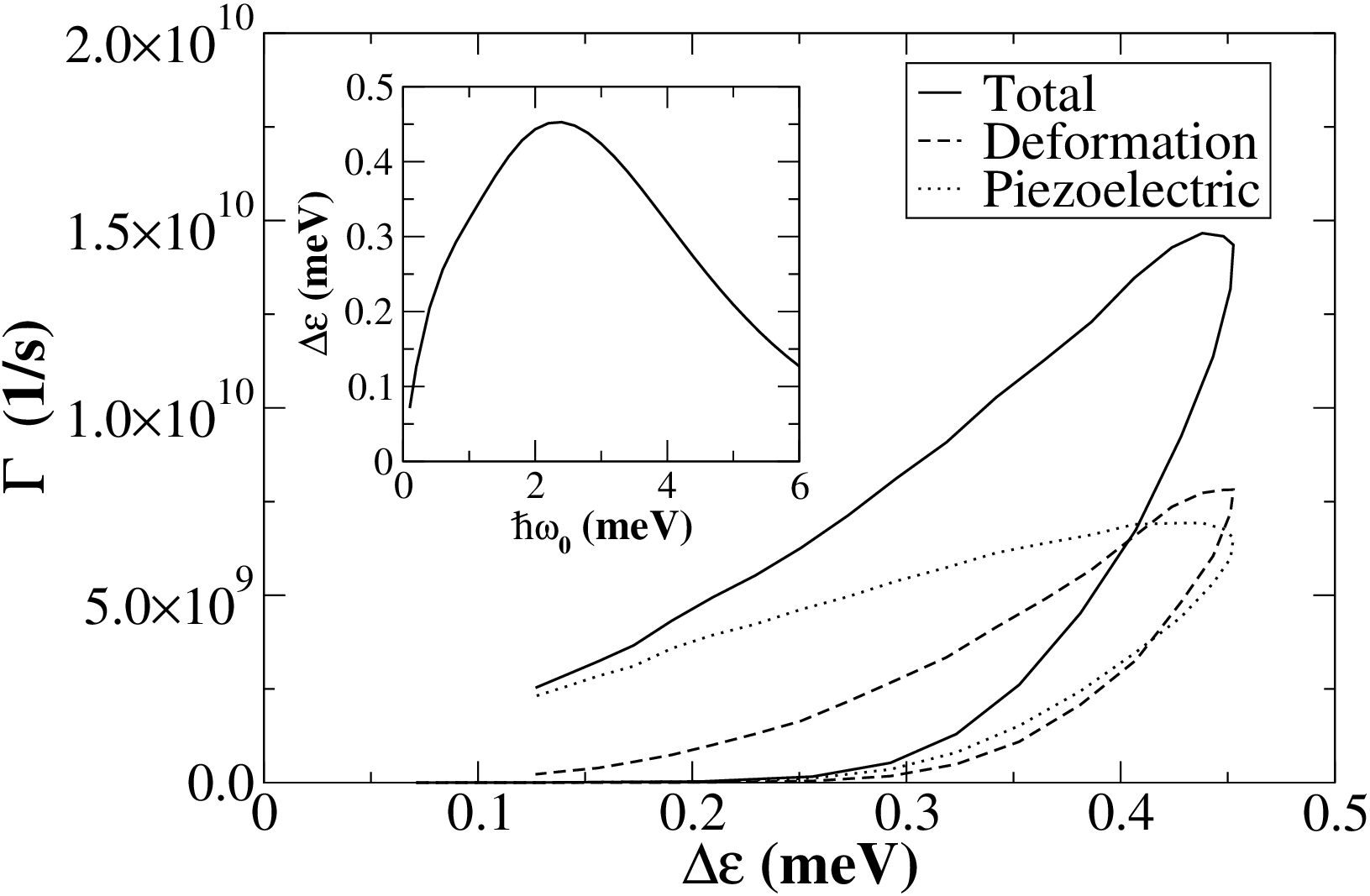}
\end{center}
\protect\caption{Electron relaxation rates as a function of the energy
splitting between the first excited state and the ground state in a double
quantum dot.  Again, rates due to deformation potential, piezoelectric
interaction, and the total relaxation rates are represented by dashed,
dotted, and straight line respectively.  The energy splitting versus the
confinement strength is given in the inset.  The interdot distance is
$2\alpha = 60$ nm.}
\label{fig_3}
\end{figure}

In Fig.~\ref{fig_2} we plot the dependence of the relaxation rates on the
confinement strength of one of the quantum dots (they are considered to be
identical in this problem) for a fixed interdot distance of $2\alpha =
60$ nm.  The relaxation rates increase with increasing confinement strength
as the energy splitting between the first excited state and the ground state
increases (as illustrated by the inset of Fig.~\ref{fig_3}), until the
confinement strength reaches about $\hbar\omega \simeq 2.4~meV$.  Further
increase in confinement strength causes increase of interdot barrier and a
decrease in energy separation of the charge qubit states, so that relaxation
rates also decrease as a consequence of the reducing energy splitting.  The
differences between the contributions of the two different types of
electron-phonon interactions can again be interpreted in the same manner as
in Fig.~\ref{fig_1}. 
The relaxation rates can also be given as a function of the energy splitting
between the first excited state and the ground state $(\Delta\varepsilon)$,
as presented in Fig.~\ref{fig_3}.  Theoretically this graph more directly
reveals the behavior of the relaxation rates: that it decreases monotonically
with decreasing energy splitting between the initial and final states,
basically because of the fast decreasing phonon density of state.  In the
double dot situation we study here for each energy splitting there could be
two different dot configurations, as illustrated in the inset, thus there are
two branches for Fig.~\ref{fig_3}.  Notice that the energy splitting
dependence of the relaxation rates is not universal,\cite{Fujisawa} because
the electron-phonon matrix elements do sensitively depend on the form/size of
the electron wavefucntions.

\begin{figure}[ht]
\begin{center}
\includegraphics[trim= 0 0 0 -45, width=3.0in]{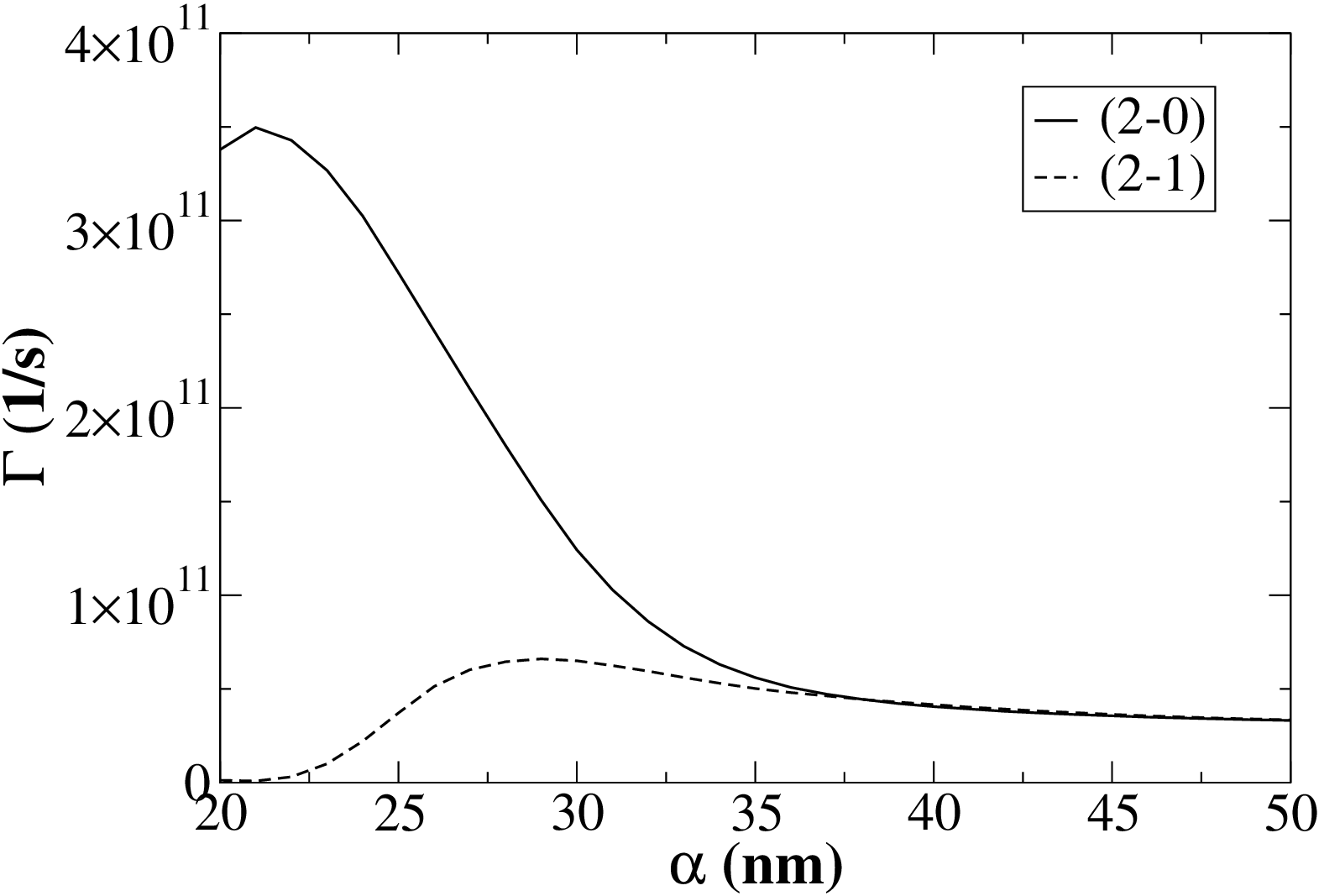}
\end{center}
\protect\caption{The relaxation rates for the second excited state as a
function of the half interdot distance $\alpha$.  The total scattering rates
include contributions from both the deformation potential and the
piezoelectric interaction.  The straight line gives relaxation rates to the
ground state, while the dashed line is the relaxation rates to the first
excited state.  The strength of the confinement is $\hbar\omega_{0}=3~meV$.}
\label{fig_4}
\end{figure}

Since we have calculated the low energy spectrum of a horizontally coupled
double quantum dot, we can easily calculate phonon emission rates when the
electron is in an excited state.  For example, Fig.~\ref{fig_4} presents the
phonon emission rates of an electron initially in the second excited state as
a function of the half interdot dot distance $\alpha$.  Now a phonon-emitting
transition can take the electron to either the first excited or the ground
state.  Furthermore, since the second excited state of a double dot is
essentially made up of the $2p$ orbitals of the two single quantum dots, the
energy splitting between the second excited and the first excited or ground
states never goes to zero: at large interdot separation, this energy
splitting approaches single electron excitation energy $\hbar \omega_0$,
which is chosen as 3 meV in this calculation.  Thus the phonon emission rate
remains finite at large interdot separation, as clearly illustrated in
Fig.~\ref{fig_4}.  In essence now electron relaxation is dominated by the
electron-phonon coupling in each of the single quantum dots.  This also
explains why the relaxation rates into the first excited state and the ground
state becomes identical as interdot separation increases. 

In summary, our results on the electron relaxation through single phonon
emission in a double quantum dot\cite{note_absorption} show a relatively
simple and straightforward dependence on energy separation of the initial and
final states (basically related to phonon density of state), and thus a
sensitive dependence on the confinement strength and interdot distance (since
energy splitting between the charge qubit states depends sensitively on these
parameters).  According to Eq.~\ref{eq:A_r} there is a strong angular
dependence for the electron-phonon coupling matrix elements.  However,
integral over the phonon wavevectors essentially averages out all the
detailed features in the final relaxation rates.  Another feature of our
results is that piezoelectric interaction dominates when the energy splitting
between the charge qubit states is small, while deformation potential
interaction dominates when the energy splitting is large.  This feature again
has a very simple physical explanation in the different ${\bf q}$-dependence
of the two types of interaction.

\subsection{Pure charge dephasing due to acoustic and optical phonons}

Electron-phonon interaction not only causes electron relaxation (or excitation
at finite temperatures) between qubit states, it can also cause dephasing
between them if the electron is in a superposition state, as we have
discussed in Section III.C.  Here we calculate the dephasing effects from
both acoustic and optical phonons.  According to Eqs.~(\ref{eq:rho_10}) and
(\ref{spectral}), the quantity $B^2(t)$ completely determines the loss of
coherence from the off-diagonal density matrix element between the charge
qubit states, therefore it is the quantity we focus on in all the results and
figures in the following.

\begin{figure}[ht]
\begin{center}
\includegraphics[trim= 0 0 0 -70, width=3.0in]{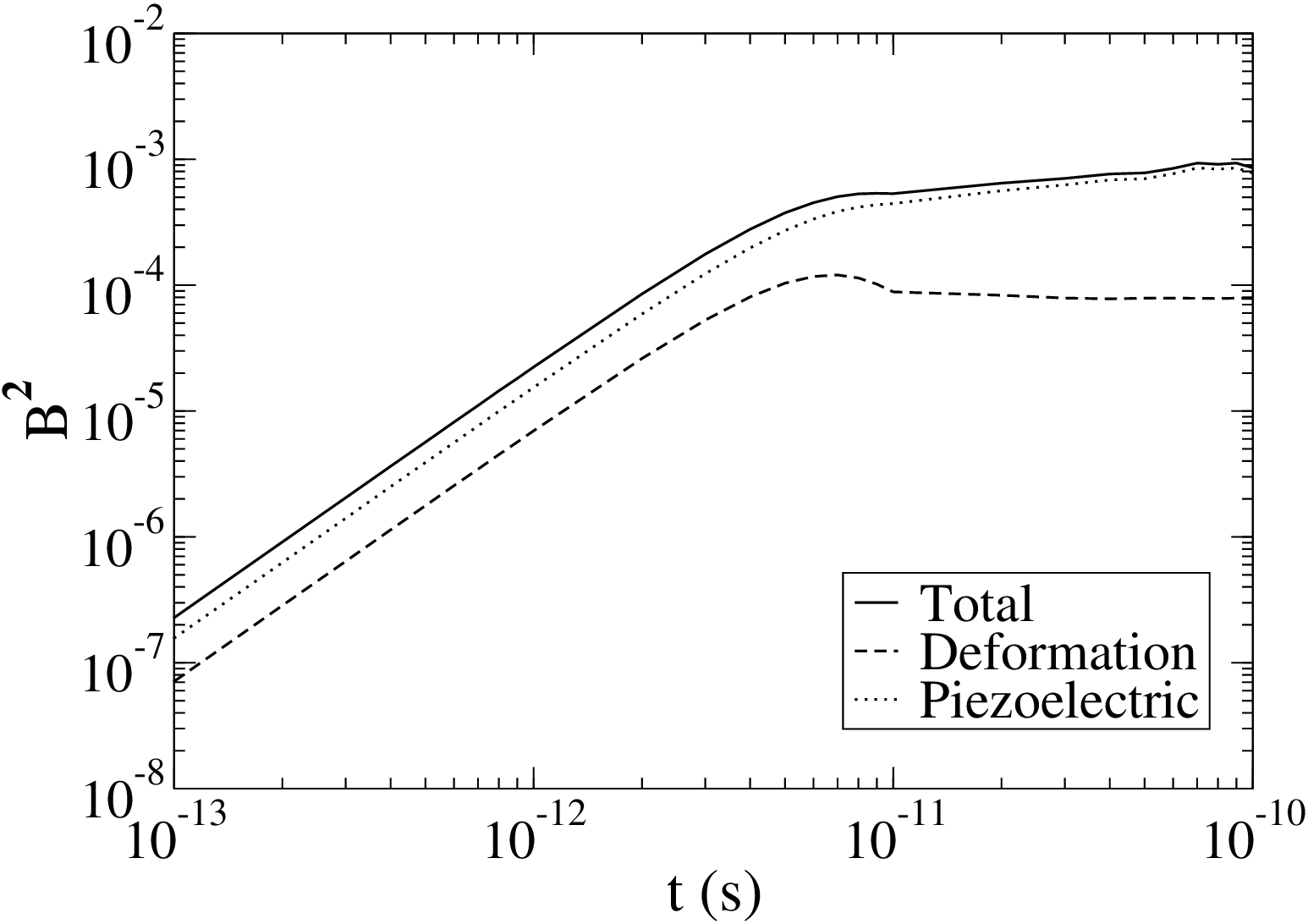}
\end{center}
\protect\caption{Dephasing rates as a function of time $t$.  The dashed,
dotted, and straight lines represented dephasing rates due to deformation
potential, piezoelectric interaction, and the total, respectively.  The
strength of the confinement $\hbar\omega_{0}=3$ meV, the interdot distance is
$2\alpha = 60$ nm, and there is no interdot bias: $V_{R} = 0$ meV.}
\label{fig_5}
\end{figure}

Figure~\ref{fig_5} presents the time dependence of dephasing between the
charge qubit states for an unbiased system due to acoustic phonons. 
According to Eq.~(\ref{eq:A_ph}), dephasing should be quite small in the
unbiased situation because it mostly comes from overlaps in between the
double dot.  An interesting feature of curves in Fig.~\ref{fig_5} is that
they rapidly increase for the first 10 ps or so, then more or less saturate,
so that $B^2(t)$ depends only very slowly on time after 100 ps. 
Mathematically, the very fast time-dependence of dephasing is due to the
trigonometric dependence on phonon frequencies and time $\sin \omega_{s}t/2$. 
As indicated in Eq.~(\ref{eq:A_ph}), only zone-center phonons contribute
significantly to dephasing (because of the integral over $\exp(i{\bf q}\cdot
{\bf r})$), with frequency ranging from zero up to $\sim \hbar c_s /\alpha$,
which is in the order of 0.1-1 THz for GaAs.  As time evolves starting from
zero, the zone-center phonon contributions quickly mix and the initial rise
of $B^2(t)$ is mostly determined by the higher frequency phonons because
their density of state is much higher.  After the initial rise, phonons with
different frequencies will not contribute always constructively, thus
producing the much flatter behavior of $B^2(t)$.  Furthermore, $B^2(t)$ will
not rise monotonically after the initial rise because the phonons contribute
through a sinusoidal function.  The time evolution shows that dephasing
quickly rises but then saturates after about $100$ ps.  

The dephasing behavior here is quite different from the ordinarily assumed
$\exp(-\gamma_ph t)$ type of behavior.  This difference is closely related to
the spin-boson type of coupling in the present problem\cite{Palma96,Duan} and
to a degree the phonon density of state of the semiconductor structure. 
There are two important consequences for the temporal behavior of $B^2(t)$. 
First, there is a very fast initial dephasing, occurring in a time period
smaller than 100 ps, due to the interaction between the qubit electron and
the acoustic phonon bath.  Second, the time-dependence of $B^2$ at large time
is very flat---it can basically be taken as a constant after 100 ps.  A
constant dephasing factor will not produce a decaying signal in terms of, for
example, oscillations in electrons.  Instead, it simply reduces the contrast
in the charge oscillation.  This can be seen easily from
Eq.~\ref{density_matrix}.  The presence of a constant $\exp(-B^2)
\sim \exp(-0.05)$ simply reduces the magnitude of $\rho_{01}$ by a constant
factor of 0.05, which is not a particularly large suppression (though
significant in terms of fault tolerant quantum computing).

\begin{figure}[ht]
\begin{center}
\includegraphics[trim= 0 0 0 -45, width=3.0in]{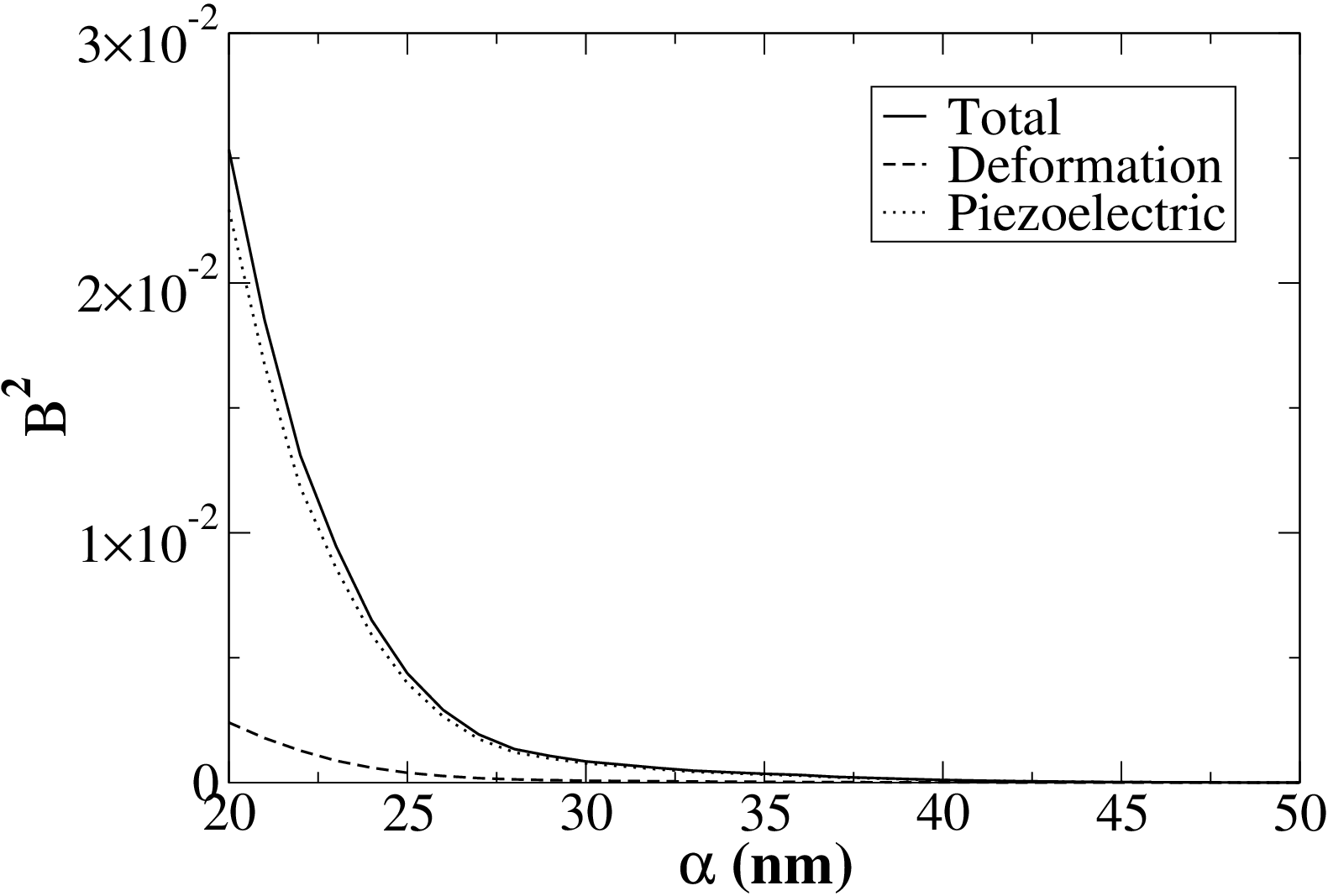}
\end{center}
\protect\caption{Dephasing factor $B^2(t)$ as a function of the half interdot
distance $\alpha$.  No bias is applied across the two quantum dots, and time
$t$ is chosen to be 60 ps.  The dephasing rates due to deformation potential,
piezoelectric interaction, and total dephasing rates are represented by
dashed, dotted, and straight line respectively.  The strength of the
confinement is $\hbar\omega_{0}=3$ meV.}
\label{fig_6}
\end{figure}

In Fig.~\ref{fig_6} we further explore the dephasing rates as a function of
the interdot distance for an unbiased system.  As indicated in
Eq.~(\ref{eq:A_ph}), if the two quantum dot are well separated, the overlap
integrals go down quickly, so that dephasing should also be strongly
suppressed.  This is exactly the behavior we observe in Fig.~\ref{fig_6}.

\begin{figure}[ht]
\begin{center}
\includegraphics[trim= 0 0 0 -45, width=3.0in]{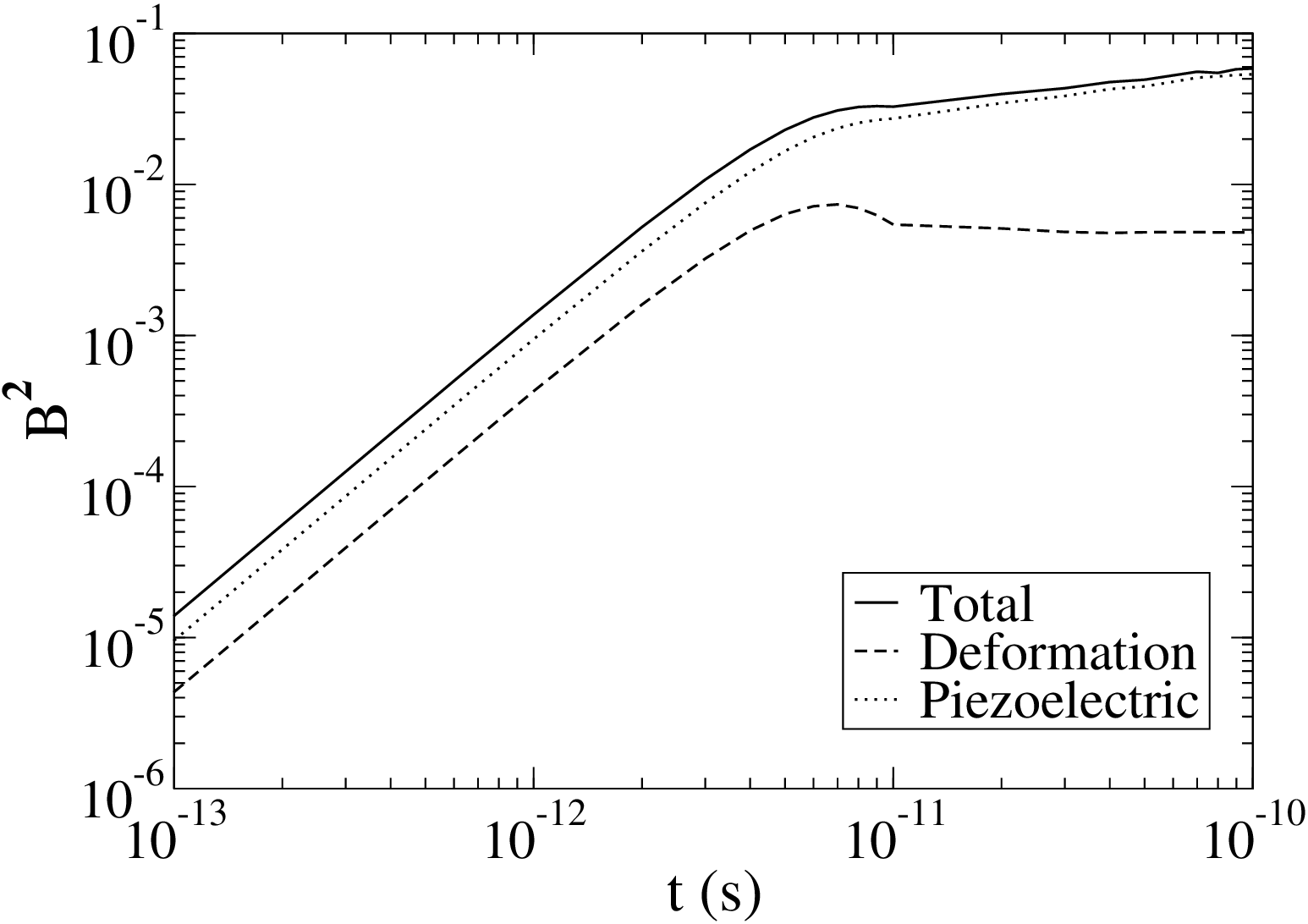}
\end{center}
\protect\caption{Dephasing rates as a function of time $t$ in the presence of
interdot bias.  Again, the dephasing rates due to deformation potential,
piezoelectric interaction, and the total rates are represented by dashed,
dotted, and straight line respectively.  The strength of the confinement is
$\hbar\omega_{0}=3$ meV, the interdot distance is $2\alpha = 60$ nm, and the
interdot bias voltage is $V_{R}=1.5$ meV.}
\label{fig_7}
\end{figure}

\begin{figure}[ht]
\begin{center}
\includegraphics[trim= 0 0 0 -45, width=3.0in]{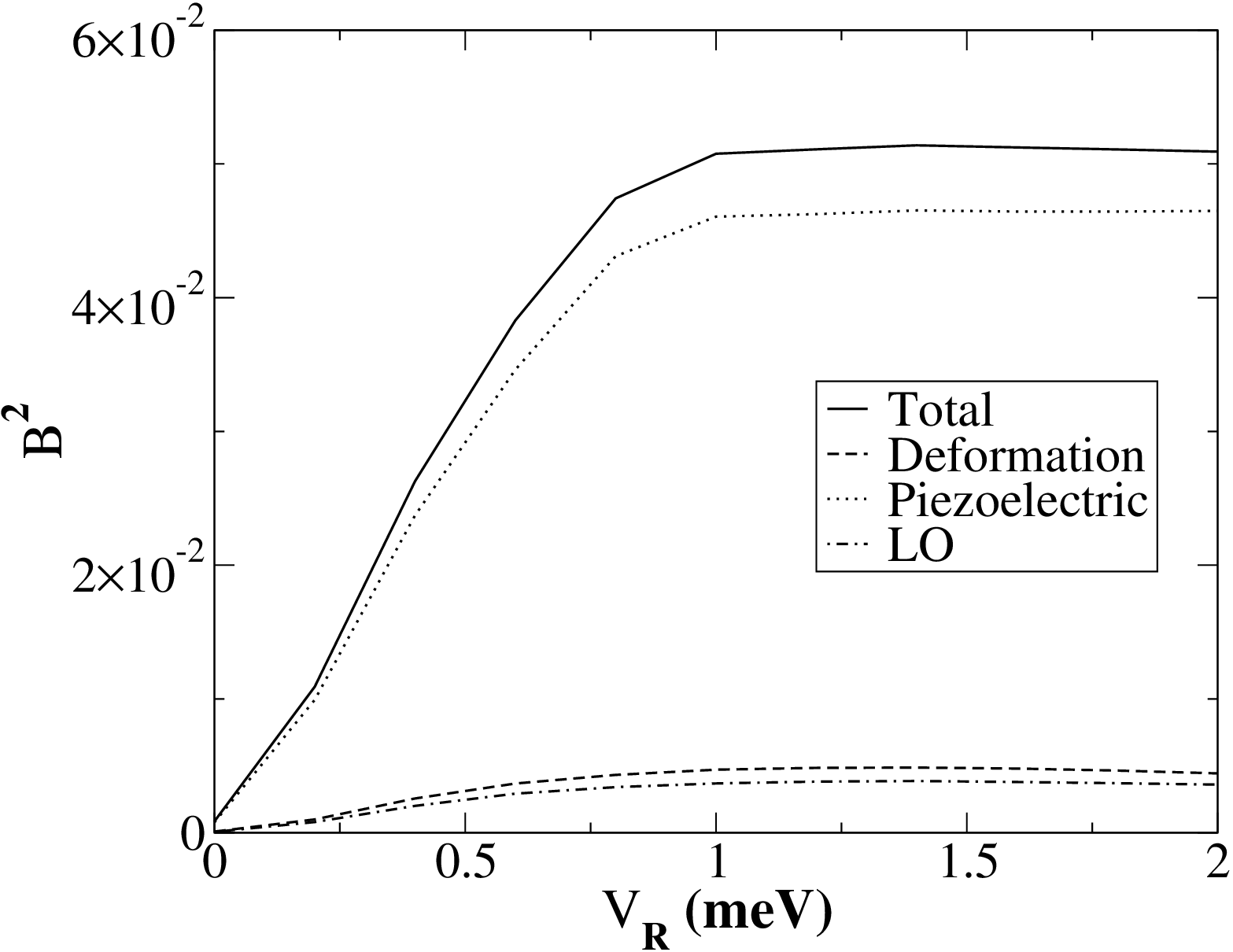}
\end{center}
\protect\caption{Dephasing rates as a function of the interdot bias voltage
$V_{R}$.  The solid line represents dephasing rates due to
electron-acoustic-phonon interaction through both deformation potential and
piezoelectric interaction.  The dashed and dotted lines represent the
deformation potential and piezoelectric contributions separately.  The
dot-dashed line represents the dephasing effects from polar interaction with
optical phonons.   Here the strength of the confinement is
$\hbar\omega_{0}=3$ meV, the interdot distance is $2\alpha = 60$ nm, and the
time of observation for the dephasing effect is $t = 60$ ps.}
\label{fig_8}
\end{figure}

When a bias is applied between the two quantum dots of the double dot, the
magnitude of dephasing should increase, which is exactly what our numerical
results presented in Figs.~\ref{fig_7} and \ref{fig_8} follow. 
Figure~\ref{fig_7} shows a very similar temporal behavior as that in the
unbiased cases, albeit with a much larger saturated value for $B^2$. 
Figure~\ref{fig_8} shows the dependence on the interdot bias voltage $V_{R}$
by the dephasing rate.  As expected from Eq.~(\ref{eq:A_ph}), the dephasing
rates increases but then saturates as the qubit states at high bias are
essentially the two single dot ground states (assuming higher excited states
still have not influenced the charge qubit states yet).  In Fig.~\ref{fig_8}
we have also included the contribution of optical phonons to dephasing. 
Although this contribution is smaller than the acoustic phonon effect, it is
still a considerable contribution.  

As we have pointed out before, optical phonons do not contribute significantly
to electron relaxation.  However, the zone-center longitudinal optical
phonons do contribute to dephasing through the polar interaction.  To better
understand the pure dephasing due to the electron polar interaction with
optical phonons, we can perform an analytical assessment for a simple double
dot configuration.  For optical phonons, $\omega_{\bf q} = \omega_{LO}$
is a constant near the zone center.  Thus at zero temperature
\begin{eqnarray}
B^{2}(t) & = & \frac{M^2}{\pi^{3} (\hbar \omega_{LO})^2} \sin^{2}
\frac{\omega_{LO} t}{2} \int d^3{\bf q} \frac{|\mathcal{I}({\bf
q})|^2}{q^2} \nonumber \\
& = & \frac{2e^2}{\pi^2 \hbar \omega_{LO}} \left( \frac{1}{\epsilon_\infty} -
\frac{1}{\epsilon_0} \right) \sin^2 \frac{\omega_{LO} t}{2} \int dq d\Omega
|\mathcal{I}({\bf q})|^2
\end{eqnarray}
The $q$-integral is now just a number that is inversely proportional to the
size of the double dot.  The magnitude of $B^2$ is thus determined by the
polar interaction strength and the double dot size, and the time-dependence of
$B^2$ is all in the sinusoidal factor.  Therefore, pure dephasing due to
optical phonons sets in at a very small time scale, in the order of 100
femtoseconds because of the fact that $\hbar \omega_{LO} \sim 36$ meV.  The
magnitude of $B^2$ is not vanishingly small, either.  Using nominally GaAs
parameters and assume spherical quantum dots with Gaussian wavefunctions
$\sim \exp(- r^2 /2a^2)$, we can estimate the order of magnitude of the
dephasing factor as
\begin{eqnarray}
B^{2}(t) & \approx & \frac{4\sqrt{2} (e^2/a)}{\sqrt{\pi} \hbar \omega_{LO}}
\left( \frac{1}{\epsilon_\infty} - \frac{1}{\epsilon_0} \right) \sin^2
\frac{\omega_{LO} t}{2} \nonumber \\
& \sim & 0.05 \sin^2 \frac{\omega_{LO} t}{2} \,.
\end{eqnarray}
For the last step we assume a wavefunction size $a \sim 20$ nm,
which corresponds to a pretty small quantum dot.  Since $B^2$ is inversely
proportional to the size $a$ of the quantum dot wavefunction, larger quantum
dot would produce a smaller dephasing magnitude for $B^2$.  Anyway, it is
clear that optical phonons produce a dephasing effect that has a comparable
if somewhat smaller magnitude as the acoustic phonons.

The dephasing effect from optical phonons evolves extremely fast, so
that the only observable effect would be its average over time, which is a
constant.  This is quite similar to the pure dephasing effect from the
acoustic phonons, which also rises rapidly ($\sim 100$ ps, slower than
optical phonons but still much faster than the ordinary time scale of
nanosecond for charge dynamics).  Recall that a constant dephasing factor
will only reduce the contrast in the measurable quantities such as charge
oscillation.  For optical phonons this amounts to a reduction in the
magnitude of $\rho_{01}$ by a constant factor of 0.95, which, like for
acoustic phonons, is not a large suppression.  Experimental techniques have
not developed to the degree to allow detection of such a small suppression of
signals.  

In short, dephasing effects on the electron orbital degrees of freedom from
electron-phonon interaction in a double dot should reveal itself mostly
through a reduction of contrast in measurable physical quantities (such as
electron oscillation between the double dot), but not a temporally decaying
signal.  Decays observed in experiments such as \onlinecite{Hayashi} should
originate from relaxation, not dephasing, if it is dominated by
electron-phonon interaction.

%
%********************
\section{CONCLUSIONS}
%********************
%

In this study we have investigated electron decoherence in a double quantum
dot due to electron-phonon coupling.  In particular, we have evaluated
electron relaxation through emission of a single acoustic phonon.  We found a
sensitive dependence of the relaxation rates on system parameters such as
confinement strength and interdot distance.  We have also evaluated electron
dephasing through interaction with both acoustic and optical
phonons---because of the absence of energy conservation requirement, all
phonon modes contribute to dephasing.

%**********************
\section{ACKNOWLEDGMENT}
%**********************
%
The work is supported in part by NSA and ARDA under ARO contract
No.~DAAD19-03-1-0128.

%**************************

%
%
%\newpage
%

\begin{thebibliography}{99}
\bibitem{bulk_phonons} In the present study we have only included the
three-dimensional bulk acoustic phonons, but not two-dimensional surface
acoustic waves\cite{Shilton,Fujisawa} or confined acoustic
waves.\cite{Stroscio,Mitin}
\bibitem{note_absorption} Our calculations presented in this paper are all
done at $T_{lat}=0$, when phonon absorption can be neglected.  At finite
lattice temperature, phonon absorption becomes finite.  However, its behavior
will not be significantly different from emission as it will also be
dominated by zone center phonons and limited by the phonon density of states.
\end{thebibliography}
\end{document}